\documentclass[preprint,showpacs,preprintnumbers,amsmath,amssymb,aps]{revtex4}

\usepackage{graphicx}% Include figure files
\usepackage{dcolumn}% Align table columns on decimal point
\usepackage{bm}% bold math

\begin{document}

\preprint{PRESAT-8101}

\title{Even-odd oscillation of conductance of 5{\it d} metal atomic nanowires}

\author{Tomoya Ono}
\email{ono@prec.eng.osaka-u.ac.jp}
\affiliation{Department of Precision Science and Technology, Osaka University, Suita, Osaka 565-0871, Japan}

\date{\today}% It is always \today, today,

\begin{abstract}
The electron-transport properties of single-row monoatomic nanowires made of 5$d$ elements are examined by first-principles calculations based on the density functional theory. We found that oscillation patterns with a period longer than two-atom length are dominant in the conductance of Ir and Pt monoatomic nanowires, although the transmission of the $s$-$d_{z^2}$ channel still exhibits even-odd oscillatory behavior. When the nanowires are deformed into zigzag configurations from the straight configuration, the oscillation behavior of the patterns with long periods changes and the oscillations eventually disappear. On the other hand, the even-odd oscillatory behavior of the $s$-$d_{z^2}$ channel still survives even in the deformed nanowire. The even-odd oscillation in the conductance of Ir and Pt nanowires is interpreted to be due to the low sensitivity of the oscillation of the $s$-$d_{z^2}$ channel to the spatial deformation of the nanowires.
\end{abstract}

\maketitle
\section{Introduction}
Metallic nanowires consisting of several atoms have been generated using scanning tunneling microscopes or mechanically controllable break junctions \cite{datta}. Since they have the simplest possible structure and represent the ultimate limit of the miniaturization of electronic devices, many theoretical and experimental studies on their electron-transport properties have been carried out. In such minute systems, the electron transport becomes ballistic and their conductance is quantized in the unit of $G_0$ (=2$e^2/h$), in contrast to the diffusive transport in macroscopic systems, where $e$ is the electron charge and $h$ is Planck's constant. In addition, several theoretical studies report oscillatory behavior with a period of two-atom length in the conductance of single-row nanowires made of monovalent atoms such as Na, K, and Au \cite{lang,emberly,nkobayashi2,nkobayashi,tsukamoto,sim,lee,egami_jim,khomyakov,egami}, which is known as even-odd oscillation.

In 2003, Smit {\it et al.} \cite{smit} experimentally found oscillatory behavior with a period of two-atom length in the conductance of Ir and Pt monoatomic nanowires in addition to that of the Au nanowire. The following year, de la Vega {\it et al.} \cite{vega} examined the electron-transport properties of Ir, Pt, and Au nanowires by a tight-binding calculation and claimed that additional oscillation patterns with a longer period and larger amplitude than those obtained in Ref. \cite{smit} can be found in the conductance of Ir and Pt nanowires. Although it is intuitively expected that patterns with a large amplitude are dominant in the conductance trace obtained by experiments, there have been no experimental studies reporting the existence of such oscillation patterns. From the theoretical point of view, oscillations with a period longer than two-atom length can be observed because the oscillatory behavior of the conductance is caused by the quantum-mechanical wave character of the electrons. Thygesen and Jacobsen \cite{thygesen} found that the conductance of the Al wires oscillates with a period of four atoms as the length of the wire is varied in 2003, and Garc\'\i a-Su\'arez {\it et al.} \cite{garcia} computed the even-odd oscillation in the conductance of Pt nanowires with three-, four-, and five-atom length by first-principles simulations in 2005,. However, they did not discuss the origin of the even-odd oscillation in the conductance of 5{\it d} metal atomic nanowires and the existence of the additional oscillation patterns.

In this paper, we implement a first-principles electron-transport calculation for single-row Ir, Pt, and Au monoatomic nanowires within the framework of the density functional theory \cite{dft}. The aims of this study are to explore whether the amplitudes of the oscillation patterns with a long period are large when they are computed by first-principles calculation and to reveal why the even-odd oscillation is dominant in the experimentally obtained conductance traces. Our results indicate that the oscillation patterns with a long period exhibit a large amplitude of {\it ca.} 0.5 $G_0$. The even-odd oscillation is insensitive to the structural deformation of the nanowire, while oscillations with a longer period than two-atom length are easily affected by structural deformation. This implies that only the even-odd oscillation can survive and the other oscillation patterns are cancelled out by being averaged over the thousands of scans in the experiment \cite{smit}.

The rest of this paper is organized as follows. In Sec. II, we briefly describe the method used in this study. Our results are presented and discussed in Sec. III. We summarize our findings in Sec. IV.

\section{Computational methods}
Our first-principles calculation method for obtaining the electron-transport properties is based on the real-space finite-difference approach \cite{rsfd,icp,tsdg}, which enables us to determine the self-consistent electronic ground state with a high degree of accuracy by a timesaving double-grid technique \cite{icp,tsdg} and the direct minimization of the energy functional \cite{dm}. Moreover, the real-space calculations eliminate the serious drawbacks of the conventional plane-wave approach, such as its inability to describe nonperiodic systems accurately. We examine the electron-transport properties of the nanowires suspended between semi-infinite electrodes by computing the scattering wave functions continuing from one electrode to the other.

The scattering wave functions from the left electrode are written as
\begin{equation*}
\Psi_i^L = \left \{
\begin{array}{ll}
\Phi_i^{in} + \sum_j r_{ij}^L \Phi_j^{ref} & \mbox{(in the left electrode)} \\
\phi_i^L & \mbox{(in the scattering region)} \\
\sum_j t_{ij}^L \Phi_j^{tra} & \mbox{(in the right electrode),}
\end{array}
\right.
\end{equation*}
where $\Phi$'s are the bulk wave functions inside the electrodes. Scattering wave functions from the right electrode can be written in a similar way. The reflection coefficients $r^L$, transmission coefficients $t^L$, and the wave function in the scattering region $\phi^L$ are evaluated by the overbridging boundary-matching formula \cite{icp,obm} under the semi-infinite boundary condition in the $z$ direction. The conductance of the nanowire system at the limits of zero temperature and zero bias is described by the Landauer-B\"uttiker formula, G=tr({\bf T}$^\dag${\bf T}) G$_0$ \cite{buttiker}, where {\bf T} is the transmission matrix. To investigate the states actually contributing to electron transport, the eigenchannels are computed by diagonalizing the Hermitian matrix {\bf T}$^\dag${\bf T} \cite{nkobayashi}.

The norm-conserving pseudopotentials \cite{norm} of Troullier and Martins \cite{tmpp} are employed to describe the electron-ion interaction, and exchange correlation effects are treated by a local density approximation \cite{lda}.

\section{Results and discussion}
We first calculate the electronic structures of infinite straight Ir, Pt, and Au wires of equal interatomic distance $d(=a_0/\sqrt{2})$, where $a_0$ is the lattice constant of fcc bulk and is chosen to be 7.25 a.u., 7.43 a.u., and 7.71 a.u. for Ir, Pt, and Au, respectively. The computational conditions are as follows: the grid spacing $h$ is set to $\sqrt{3}a_0$/34 and a denser grid spacing of $h$/3 is employed in the vicinity of nuclei by the augmentation of the double-grid technique \cite{icp,tsdg}. The supercell contains an atom, and the size of the supercell is $L_x=L_y=46h$ and $L_z=d$, where $L_x$ and $L_y$ are the lengths of the supercell in the $x$ and $y$ directions perpendicular to the wire, respectively, and $L_z$ is the length in the $z$ direction. The integration over the Brillouin zone along the wire-axis direction is performed by the equidistant sampling of 80 {\it k}-points. We ensured that the increase in the number of grid points and the number of $k$-points sampled did not affect our conclusion. 

\begin{figure*}[htb]
\begin{center}
\includegraphics{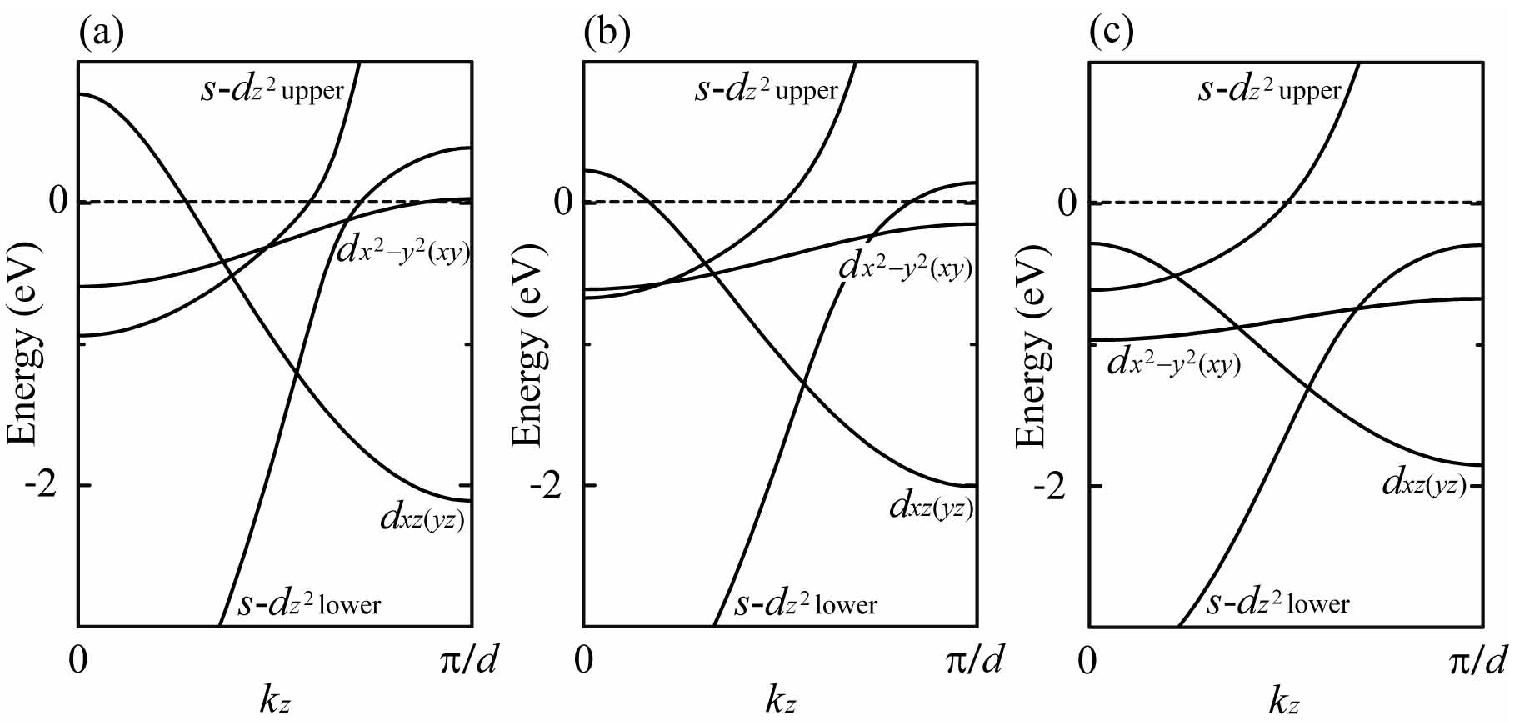}
\end{center}
\caption{Energy band structures of infinite (a) Ir, (b) Pt, and (c) Au nanowires. Zero of energy is chosen to be the Fermi level.}
\label{fig:inf-band}
\end{figure*}

\begin{figure}[htb]
\begin{center}
\includegraphics{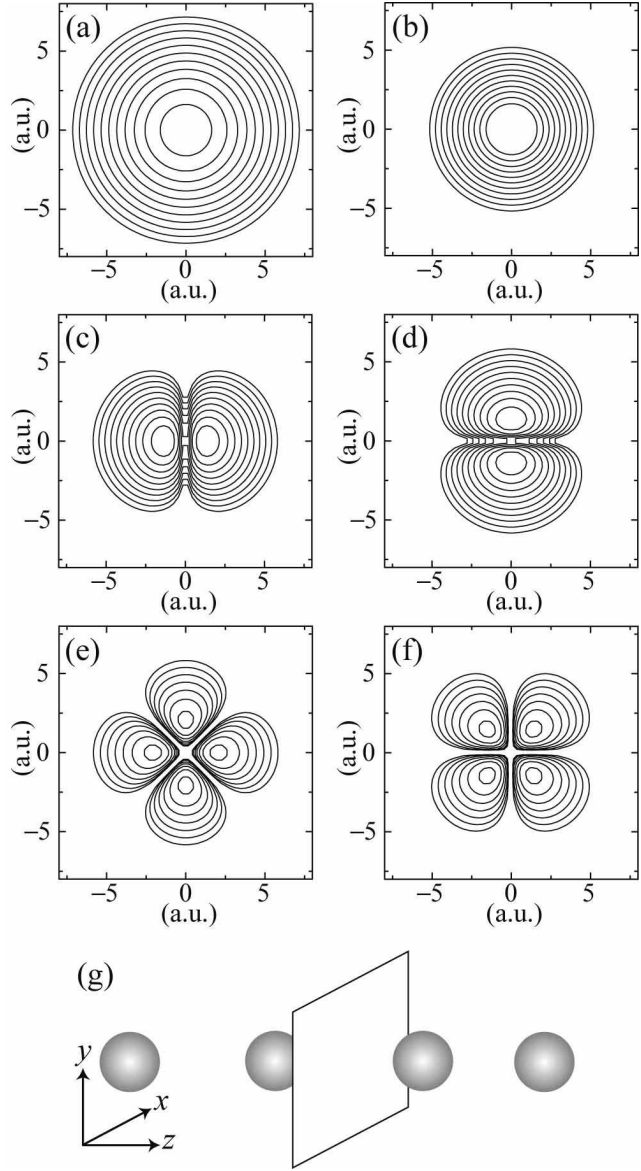}
\end{center}
\caption{Charge density distributions of (a) $s$-$d_{z^2}$ upper, (b) $s$-$d_{z^2}$ lower, (c) $d_{xz}$, (d) $d_{yz}$, (e) $d_{x^2-y^2}$, and (f) $d_{xy}$ bands of infinite Au wire. Each contour represents twice or half the density of the adjacent contour lines. The lowest contour line corresponds to 7.81 $\times 10^{-6}$ (electrons/bohr$^3$). (g) Schematic description of the plane where the distributions are shown. The distributions of the infinite Ir and Pt wires are almost same as those of the Au wire.}
\label{fig:inf-density_distribution}
\end{figure}

Figures~\ref{fig:inf-band} and \ref{fig:inf-density_distribution} respectively show the energy band structures of the infinite Ir, Pt, and Au wires and the charge density distributions of the valence electrons of the infinite Au wire. The bands are decomposed into atomic orbitals and labeled according to the atomic orbitals mainly constituting the bands. The charge distributions of the $d_{xz}$, $d_{yz}$, $d_{x^2-y^2}$, and $d_{xy}$ bands are divided into two or four bunches on the plane perpendicular to the wire by the nodes of the wave functions and look to be two or four thin wires. Only the upper $s$-$d_{z^2}$ band gets across the Fermi level in the case of the Au wire, while the other $d$ bands also cross the Fermi level in the case of the Ir and Pt wires. These results are consistent with the experimental results in which the maximum conductance of the Au nanowire is 1 $G_0$ while the conductances of Ir and Pt nanowires exceed 1 $G_0$ \cite{smit}.

\begin{figure}[htb]
\begin{center}
\includegraphics{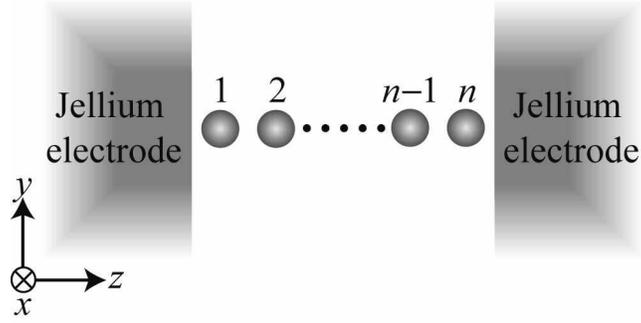}
\end{center}
\caption{Schematic description of the scattering region of monoatomic nanowire suspended between semi-infinite jellium electrodes.}
\label{fig:atomistic-model}
\end{figure}

\begin{figure}[htb]
\begin{center}
\includegraphics{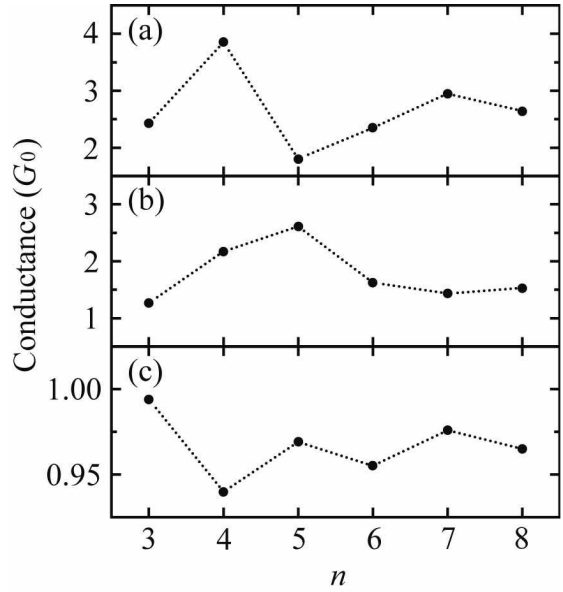}
\end{center}
\caption{Conductance of (a) Ir, (b) Pt, and (c) Au nanowires as a function of the number of atoms constituting the nanowires $n$.}
\label{fig:conductance}
\end{figure}

We next examine the electron-transport properties of the nanowires suspended between semi-infinite electrodes. Figure~\ref{fig:atomistic-model} shows the computational model. We adopt a model in which the nanowires are directly attached to structureless jellium electrodes so as to eliminate the complex effect arising from the atomic configurations of the electrodes and to reveal the effects on the electron-transport properties that are intrinsic to the atomic wire. Straight Ir, Pt, and Au nanowires of equal interatomic distance $d$ are suspended between the jellium electrodes. The distance between the edge atom of the nanowire and the surface of the jellium electrode is 2$a_0/\sqrt{6}$. The Wigner-Seitz radii of the jellium electrodes are taken to be $r_s=1.36$ a.u., $r_s=1.35$ a.u., and $r_s=1.35$ a.u. for the Ir-, Pt-, and Au-nanowire models, respectively, so that the elements of the electrodes correspond to those of the nanowires. To determine the Kohn-Sham effective potential, a conventional supercell is employed under the periodic boundary condition in all directions; the length of the supercell in the $z$ direction is $L_z=nd+80h$. The other computational conditions are the same as those in the case of the infinite wires. We compute the electron-transport properties of the nanowires using the procedure described in the previous section by imposing semi-infinite boundary conditions at both ends of the electrodes. Figure~\ref{fig:conductance} shows the conductance of the Ir, Pt, and Au nanowires. Tables~\ref{tbl:tbllir},~\ref{tbl:tbllpt}, and~\ref{tbl:tbllau} show the transmissions of the eigenchannels of the Ir, Pt, and Au nanowires, respectively. The quantum numbers of the eigenchannels that correspond to those in the case of the infinite wires are assigned according to their charge density distributions. The conductance of the Au nanowire exhibits even-odd oscillation with an amplitude of $\sim$0.02, which agrees with the experimental result \cite{smit}. On the other hand, for both Ir and Pt nanowires, although oscillatory behavior is not observed in the conductance, even-odd oscillation emerges in the transmission of the upper $s$-$d_{z^2}$ channel. In addition, the other channels manifest oscillation with a longer period and a larger amplitude of {\it ca.} 0.5 $G_0$ in agreement with the result obtained by the tight-binding calculation \cite{vega}. Moreover, the long period of the oscillation of the $d_{x^2-y^2}$ and $d_{xy}$ channels in the Ir nanowire is consistent with the conclusion of our previous study \cite{EgamiJPC}, in which we found that the period of oscillation coincides with the least common multiple of $\pi/k_z$ and the geometric period of the nanowire, i.e., the interatomic distance.

The oscillatory behavior of the transmission is led by the quantum-mechanical wave character of the electrons \cite{egami}; since the upper $s$-$d_{z^2}$ band of the infinite wire gets across the Fermi level at $k_z \approx \pi/2d$, the period of the oscillation becomes $\pi$ divided by $\pi/k_z(\approx2d)$. The oscillation of the lower $s$-$d_{z^2}$ channel cannot be observed in the experiments because of its long oscillation period. However, the $d_{xz}$, $d_{yz}$, $d_{x^2-y^2}$, and $d_{xy}$ channels are expected to contribute significantly to the experimentally obtained conductance trace, since these channels exhibit a rather short period ($<4d$) and their amplitudes are larger than that of the upper $s$-$d_{z^2}$ channel.

To explore the reason why the oscillations that arise from the $d_{xz}$, $d_{yz}$, $d_{x^2-y^2}$, and $d_{xy}$ channels do not appear in the experimentally obtained traces, we examine the transport properties of zigzag Pt nanowires because some theoretical calculations indicate that in the zigzag configuration, the angle subtended between the bonds and the nanowire axis is up to 30 degrees \cite{garcia,disorder}. Moreover, it is noteworthy that the experimentally obtained conductance traces are averaged over several thousand scans, during which the nanowire is not always straight due to the effect of the atomic structure of the electrodes. We move the odd- (even-) numbered atoms in Fig.~\ref{fig:atomistic-model} by $\Delta d$ ($-\Delta d$) in the $x$ direction. Tables~\ref{tbl:tblz2pt} and ~\ref{tbl:tblz5pt} show the transmissions of the channels when $\Delta d$ is 0.2 and 0.5 a.u., respectively. The angle between the bonds and the nanowire axis is 4.4 (10.0) degrees at $\Delta d$=0.2 a.u. ($\Delta d$=0.5 a.u.). Therefore, these configurations can be formed during the elongation of the nanowire.

Interestingly, the oscillatory behavior of the $s$-$d_{z^2}$ and $d_{yz}$ channels is insensitive to the spatial deviation of the nanowire. On the other hand, the transmission of the $d_{xz}$ channel of the 6-atom nanowire becomes larger than that of the 4-atom nanowire at $\Delta d$=0.2 a.u.; the phase of the oscillation of the $d_{xz}$ channel slightly moves to the longer side. At $\Delta d$=0.5 a.u., the oscillation of the $d_{xz}$ channel completely disappears. Since the plane that divides the charge density distribution of the $d_{xz}$ channel also forms a zigzag structure following the deviation of the nanowires and the distribution behaves like a two thin zigzag wire, the transmissions of the $d_{xz}$ channel are easily affected by the structural deformation of the nanowire in the $x$ direction. It is noteworthy that the oscillations of the channels whose charge density distributions are divided by the plane perpendicular to the nanowire axis are removed because the experimentally obtained conductance trace is averaged over thousands of scans. In addition, since the deviation from the straight nanowire can arise in various directions and at various positions of the nanowire, the oscillations of the $d_{yz}$ and $d_{xz}$ channels disappear and only the oscillation from the $s$-$d_{z^2}$ channel can survive. Moreover, the reduction of the transmission of the $d_{xz}$ channel at $\Delta d$=0.5 a.u. with increasing the nanowire length may contribute to the slope of the conductance trace at about (0.3$-$0.4)$G_0$/nm.

\section{Conclusion}
We have implemented a first-principles investigation of the electron-transport properties of nanowires consisting of 5$d$ atoms. Our results demonstrate that the charge density distributions of the channels are divided into two or four bunches on the plane perpendicular to the wire, with the exception of the $s$-$d_{z^2}$ channel. Oscillation patterns with a period longer than two-atom length and a larger amplitude of {\it ca.} 0.5 $G_0$ are found in the conductance of the straight Ir and Pt nanowires. These oscillations are caused by the channels whose charge density distributions are divided by the node of the wave functions. However, the transmission properties of these channels are easily affected by the structural deformation of the nanowires in the directions perpendicular to the nanowire axis. Therefore, only the even-odd oscillation of the $s$-$d_{z^2}$ channel remains in the experimentally obtained conductance traces while the oscillations of the other channels are washed out by being averaged over many scans. Although the first-principles electron-transport calculation is not a suitable tool for averaging results over numerous implementations, it is expected that the averaging effect can be confirmed by simpler tools such as tight-binding approximations.

\section*{Acknowledgements}
The author would like to thank Professor Kikuji Hirose of Osaka University for reading the entire text in its original form. This research was partially supported by a Grant-in-Aid for the 21st Century COE ``Center for Atomistic Fabrication Technology'', by a Grant-in-Aid for Scientific Research in Priority Areas ``Development of New Quantum Simulators and Quantum Design'' (Grant No. 17064012), and also by a Grant-in-Aid for Young Scientists (B) (Grant No. 17710074) from the Ministry of Education, Culture, Sports, Science and Technology. The numerical calculation was carried out using the computer facilities at the Institute for Solid State Physics at the University of Tokyo, the Research Center for Computational Science at the National Institute of Natural Science, and the Information Synergy Center at Tohoku University.

\begin{table}[thbp]
\begin{center}
\caption{Channel transmissions of straight Ir nanowire.}
\label{tbl:tbllir}
\begin{tabular}{c|cccc} \hline\hline
\hspace{3mm}$n$\hspace{3mm} & $s$-$d_{z^2}$ upper \hspace{1mm} & \hspace{1mm} $s$-$d_{z^2}$ lower & \hspace{3mm} $d_{xz(yz)}$ \hspace{3mm} & \hspace{3mm} $d_{x^2-y^2(xy)}$ \hspace{3mm} \\ \hline
3 & 0.854 & 0.016 & 0.763 & 0.016 \\
4 & 0.973 & 0.011 & 0.935 & 0.497 \\
5 & 0.727 & 0.018 & 0.510 & 0.019 \\
6 & 1.000 & 0.020 & 0.656 & 0.010 \\
7 & 0.838 & 0.056 & 0.979 & 0.048 \\
8 & 0.975 & 0.116 & 0.569 & 0.206 \\ \hline\hline
\end{tabular}
\\
\end{center}
\end{table}

\begin{table}[thbp]
\begin{center}
\caption{Channel transmissions of straight Pt nanowire.}
\label{tbl:tbllpt}
\begin{tabular}{c|cccc} \hline\hline
\hspace{3mm}$n$\hspace{3mm} & $s$-$d_{z^2}$ upper \hspace{1mm} & \hspace{1mm} $s$-$d_{z^2}$ lower & \hspace{3mm} $d_{xz(yz)}$ \hspace{3mm} & \hspace{3mm} $d_{x^2-y^2(xy)}$ \hspace{3mm} \\ \hline
3 & 0.989 & 0.000 & 0.139 & 0.000 \\
4 & 0.994 & 0.000 & 0.589 & 0.000 \\
5 & 0.984 & 0.000 & 0.814 & 0.000 \\
6 & 0.999 & 0.000 & 0.314 & 0.000 \\
7 & 0.985 & 0.000 & 0.226 & 0.000 \\
8 & 0.999 & 0.000 & 0.266 & 0.000 \\ \hline\hline
\end{tabular}
\\
\end{center}
\end{table}

\begin{table}[thbp]
\begin{center}
\caption{Channel transmissions of straight Au nanowire.}
\label{tbl:tbllau}
\begin{tabular}{c|cccc} \hline\hline
\hspace{3mm}$n$\hspace{3mm} & $s$-$d_{z^2}$ upper \hspace{1mm} & \hspace{1mm} $s$-$d_{z^2}$ lower & \hspace{3mm} $d_{xz(yz)}$ \hspace{3mm} & \hspace{3mm} $d_{x^2-y^2(xy)}$ \hspace{3mm} \\ \hline
3 & 0.994 & 0.000 & 0.011 & 0.000 \\
4 & 0.940 & 0.000 & 0.001 & 0.000 \\
5 & 0.969 & 0.000 & 0.000 & 0.000 \\
6 & 0.955 & 0.000 & 0.000 & 0.000 \\
7 & 0.976 & 0.000 & 0.000 & 0.000 \\
8 & 0.965 & 0.000 & 0.000 & 0.000 \\ \hline\hline
\end{tabular}
\\
\end{center}
\end{table}

\begin{table*}[thbp]
\begin{center}
\caption{Channel transmissions of zigzag Pt nanowire when $\Delta d=$0.2 a.u.}
\label{tbl:tblz2pt}
\begin{tabular}{c|ccccc} \hline\hline
\hspace{3mm}$n$\hspace{3mm} & $s$-$d_{z^2}$ upper \hspace{1mm} & \hspace{1mm} $s$-$d_{z^2}$ lower & \hspace{3mm} $d_{xz}$ \hspace{3mm} & \hspace{3mm} $d_{yz}$ \hspace{3mm} & \hspace{3mm} $d_{x^2-y^2(xy)}$ \hspace{3mm} \\ \hline
3 & 0.982 & 0.000 & 0.118 & 0.132 & 0.000 \\
4 & 0.985 & 0.000 & 0.361 & 0.563 & 0.000 \\
5 & 0.971 & 0.000 & 0.985 & 0.812 & 0.000 \\
6 & 0.997 & 0.000 & 0.375 & 0.311 & 0.000 \\
7 & 0.990 & 0.000 & 0.230 & 0.224 & 0.000 \\
8 & 1.000 & 0.000 & 0.239 & 0.280 & 0.000 \\ \hline\hline
\end{tabular}
\\
\end{center}
\end{table*}

\begin{table*}[thbp]
\begin{center}
\caption{Channel transmissions of zigzag Pt nanowire when $\Delta d=$0.5 a.u.}
\label{tbl:tblz5pt}
\begin{tabular}{c|ccccc} \hline\hline
\hspace{3mm}$n$\hspace{3mm} & $s$-$d_{z^2}$ upper \hspace{1mm} & \hspace{1mm} $s$-$d_{z^2}$ lower & \hspace{3mm} $d_{xz}$ \hspace{3mm} & \hspace{3mm} $d_{yz}$ \hspace{3mm} & \hspace{3mm} $d_{x^2-y^2(xy)}$ \hspace{3mm} \\ \hline
3 & 0.933 & 0.000 & 0.053 & 0.099 & 0.000 \\
4 & 0.967 & 0.000 & 0.026 & 0.400 & 0.000 \\
5 & 0.951 & 0.000 & 0.013 & 0.791 & 0.000 \\
6 & 1.000 & 0.000 & 0.006 & 0.250 & 0.000 \\
7 & 0.895 & 0.000 & 0.001 & 0.223 & 0.000 \\
8 & 0.998 & 0.000 & 0.001 & 0.516 & 0.000 \\ \hline\hline
\end{tabular}
\\
\end{center}
\end{table*}

\end{document}